\documentclass[prl,floats,twocolumn,showpacs,floatfix,superscriptaddress]{revtex4}
\usepackage{graphics,graphicx,dcolumn,bm,fleqn,epic,eepic,float}
\usepackage{amssymb,amsmath,multirow,rotate,color,float,mciteplus}
\usepackage[latin1]{inputenc}

\usepackage[dvips]{epsfig}

\begin{document}

\title{Electro-osmosis on anisotropic super-hydrophobic surfaces}

\author{Aleksey V. Belyaev}
\affiliation{Department of Physics, M. V. Lomonosov Moscow State University, 119991 Moscow, Russia}
\affiliation{A.N.~Frumkin Institute of Physical
Chemistry and Electrochemistry, Russian Academy of Sciences, 31
Leninsky Prospect, 119991 Moscow, Russia}

\author{Olga I. Vinogradova}
\affiliation{Department of Physics, M. V. Lomonosov Moscow State University, 119991 Moscow, Russia}
\affiliation{A.N.~Frumkin Institute of Physical
Chemistry and Electrochemistry, Russian Academy of Sciences, 31
Leninsky Prospect, 119991 Moscow, Russia}
\affiliation{ITMC and DWI, RWTH Aachen, Pauwelsstr. 8,
52056 Aachen, Germany}

\newcommand\Xsin{\mbox{sin}}
\newcommand\Xcos{\mbox{cos}}
\newcommand\Xsec{\mbox{sec}}
\newcommand\Xlog{\mbox{ln}}

\newcommand*{\mycommand}[1]{\texttt{\emph{#1}}}
\def\p{\par $\bullet$ }

\begin{abstract}

We give a general theoretical description of electro-osmotic flow  at striped super-hydrophobic surfaces in a thin double layer limit, and derive a relation between the electro-osmotic mobility and hydrodynamic slip-length tensors. Our analysis demonstrates that electro-osmotic flow shows a very rich behavior controlled by slip length and charge at the gas sectors. In case of uncharged liquid-gas interface, the flow is the same or inhibited relative to flow in homogeneous channel with zero interfacial slip. By contrast, it can be amplified by several orders of magnitude provided slip regions are uniformly charged. When gas and solid regions are oppositely charged, we predict a  flow reversal, which suggests a possibility of huge electro-osmotic slip even for electro-neutral surfaces. On the basis of these observations we suggest strategies for practical microfluidic mixing
devices. These results provide a framework for the rational design of super-hydrophobic surfaces.

\end{abstract}

\maketitle

{\bf Introduction.}--
Electro-osmotic (EO) ``plug'' flows are established when an electric
field forces the diffuse ionic cloud adjacent to a charged
surface in an electrolyte solution into motion. This classical subject of colloid science~\cite{lyklema1995} is currently experiencing a renaissance in micro- and nanofluidics~\cite{stone2004,eijkel.jct:2005}, which raises fundamental question of how to pump and mix fluids at
micron scales, where pressure-driven flows and inertial
instabilities are suppressed by viscosity. Electro-osmosis offers unique advantages in this area of research and technologies, such as low
hydrodynamic dispersion, no moving parts, electrical actuation and
sensing, energy conversion and storage, and easy integration with microelectronics.

Until recently,
almost all studies of EO have assumed uniform surface charge
 and no-slip hydrodynamic boundary conditions at the surface. In such a situation the \emph{scalar} electro-osmotic mobility $M_1$, which relates an apparent EO  ``slip'' velocity $\textbf{\emph{U}}_1$ (outside of the \emph{thin} double layer) to the tangential electric field $\textbf{\emph{E}}_t$ is given by the classical
Smoluchowski formula~\cite{anderson.jl:1989}

\begin{equation}\label{smoluchovsky}
    M_1 =-\frac{ U_1}{E_t}=\frac{ q_1}{\eta \kappa},
\end{equation}
 where $\eta$ is the viscosity of the solution, $ q_1$ is the \emph{constant} charge density of the no-slip surface, which can be related to the zeta potential across the diffuse (flowing) part of the double layer, $\zeta_1=q_1/\kappa\varepsilon$, where $\varepsilon$ is the permittivity of the solution, and $\kappa=\lambda_D^{-1}$ is the inverse Debye screening length, that characterizes the thickness of the electrical Debye layer (EDL).

Recent studies demonstrated the existence of a
hydrodynamic slip at hydrophobic smooth and homogeneous surface, which can be quantified by the slip
length $b$ (the distance within the solid at which the flow
profile extrapolates to zero)~\citep{vinogradova1999,lauga2005}.
The combination of the strategies of EO and hydrophobic slip, can yield enhanced EO flow.

For a charge density $q_2$ of the slipping interface, simple arguments show that the EO mobility is given by ~\citep{muller.vm:1986,joly2004}:
\begin{equation}\label{isotropic}
      M_2 =-\frac{ U_2}{E_t} = \frac{q_2}{\eta \kappa} (1 + b \kappa)
\end{equation}
Since the EO flow amplification scales as $(1+b\kappa)$, and $b$ can be of the order of tens of nanometers ~\citep{vinogradova:03,charlaix.e:2005,joly.l:2006,vinogradova.oi:2009}, for typically nanometric Debye length an order of magnitude enhancement might be expected.

\begin{figure}
\begin{center}
\includegraphics [width=8.5 cm]{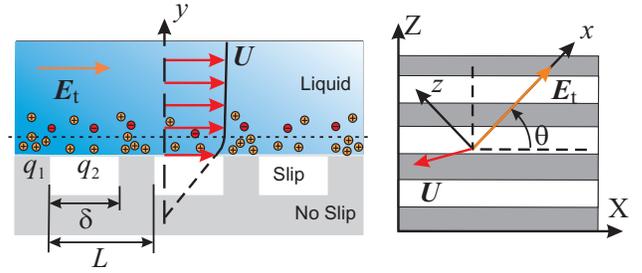}
   \end{center}
  \caption{(Color online) (a) Sketch of the superhydrophobic effective slippage effect on the EO flow. The real situation is approximated by a periodic cell of size $L$, with patterns of charges and flow boundary conditions (b) Illustration of tensorial EO response: $\theta = \pi/2$ corresponds to transverse, whereas $\theta = 0$ to longitudinal stripes.}
  \label{fig:geometry}
\end{figure}

It is now natural to assume that a massive amplification of EO flow can be reached on super-hydrophobic surfaces where effective, in general case tensorial, slip length, $\mathbf{b}_{\rm eff}$, could be the order of several microns~\citep{ou2005,joseph.p:2006,tsai.p:2009}. The controlled generation of such flows is by no means obvious, since both the slip length and the electric charge distribution on a SH surface are inhomogeneous and often anisotropic. Despite its fundamental and practical significance EO flow over SH surfaces has received little attention. Recently, \cite{Squires08} investigated EO flow past inhomogeneously charged, flat SH surfaces in the case of thick channels ($h \gg L$), thin EDL ($\kappa L \gg 1$), and predicted

\begin{equation}
{\bf M} =  M_1\cdot\left(\mathbf{I} +  \frac{q_2}{q_1} {\bf b}_{\rm eff} \kappa \right)
\label{squires}
\end{equation}
by using the Lorentz
reciprocal theorem for the Stokes flow and by assuming \emph{perfect} slip  ($b = \infty$) at gas sectors. Here $\mathbf{I}$ is the unit tensor, and  we keep notations, $q_1$ and $q_2$, to characterize the surface charge density at the no-slip and slip regions, as above.  This expression indicates negligible flow enhancement in case of an uncharged liquid-gas interface (which has been confirmed by later studies~\citep{Huang08,bahga:2009}), and shows that surface anisotropy generally leads to a \emph{tensorial} EO response.

In this Letter, a general  situation of EO flow past SH surfaces with patterns of arbitrary \emph{partial} slip, is considered (Fig.\ref{fig:geometry}). Our focus is on the canonical EO geometry of a
thick parallel-plate channel with a two-component (no-slip and slip) coarse texture, varying on
scales larger than the EDL thickness.

{\bf General theory.}-- To highlight the effect of anisotropy, we focus on an idealized, flat, periodic, charged, striped SH surface in the Cassie state, sketched in Fig.~\ref{fig:geometry}, where  the liquid-solid interface has no slip ($b_1=0$) and the liquid-gas interface has partial slip ($b_2=b,\,\, 0 \le b \le \infty $). As a simple estimate, lubricating gas sectors of height $e$ with
viscosity $\eta_g$ much smaller than
$\eta$~\cite{vinogradova.oi:1995d} have a local slip length $b_2 \approx e
(\eta/\eta_g)  \approx 50\, e$, which can reach tens of $\mu$m. Let then $\phi_1$ and $\phi_2=\delta/L$ be the area fractions of the solid and gas phases with $\phi_1+\phi_2=1$. Pressure-driven flow past such stripes has been shown to
depend on the direction of the flow, and the eigenvalues of the slip-length tensor~\cite{Bazant08} read~\cite{belyaev.av:2010a}
\begin{equation}\label{beff_par_largeH}
  b_{\rm eff}^{\parallel} \simeq \frac{L}{\pi} \frac{\ln\left[\sec\left(\displaystyle\frac{\pi \phi_2}{2 }\right)\right]}{1+\displaystyle\frac{L}{\pi b}\ln\left[\sec\displaystyle\left(\frac{\pi \phi_2}{2 }\right)+\tan\displaystyle\left(\frac{\pi \phi_2}{2}\right)\right]},
\end{equation}
\begin{equation}\label{beff_ort_largeH}
  b_{\rm eff}^{\perp} \simeq \frac{L}{2 \pi} \frac{\ln\left[\sec\left(\displaystyle\frac{\pi \phi_2}{2 }\right)\right]}{1+\displaystyle\frac{L}{2 \pi b}\ln\left[\sec\displaystyle\left(\frac{\pi \phi_2}{2 }\right)+\tan\displaystyle\left(\frac{\pi \phi_2}{2}\right)\right]}.
\end{equation}
These expressions depend strongly on a texture period $L$. When $b/L \ll 1$ they predict the area-averaged isotropic slip length, $b_{\rm eff}^{\perp, \parallel} \simeq \phi_2 b$. When $b/L \gg 1$,  expressions (\ref{beff_par_largeH}) and (\ref{beff_ort_largeH}) take form
\begin{equation}\label{beff_ort_largeH_id}
  b_{\rm eff}^{\perp} \simeq \frac{L}{2 \pi} \ln\left[\sec\left(\displaystyle\frac{\pi \phi_2}{2 }\right)\right],\quad b_{\rm eff}^{\parallel}\simeq2 b_{\rm eff}^{\perp},
\end{equation}
that coincides with results obtained for the perfect slip ($b = \infty$) stripes~\cite{lauga.e:2003}.

The EO mobility is represented by $2 \times 2$ matrices diagonalized by a rotation~\cite{bahga:2009}.
By symmetry, the eigen-directions of $\mathbf{M}$ correspond to longitudinal ($\theta=0$) and transverse ($\theta=\pi/2$) alignment with the applied electric field (Fig.~\ref{fig:geometry}), so we need only to compute the  eigenvalues, $M^{\parallel}$ and $M^\perp$, for these cases.

 We consider a semi-infinite electrolyte in the region $y>0$ above a patterned surface at $y=0$ subject to an electric field, $\textbf{\emph{E}}_t=E_t \hat{\textbf{\emph{x}}}$, in the $x$ direction. For nano-scale patterns ($L < 1\mu$m), we can neglect convection ($Pe \ll 1$ for a typical ionic diffusivity $D$), so that  $\psi(x,y,z)$ is independent of the fluid flow~\cite{epaps}. We also assume weak field ($|E_t| L \ll |\psi|$) and weakly charged surface ($|\psi|\ll k_B T/ (z e) = 25/z$ mV) for a $z:z$ electrolyte, so that $\psi$ satisfies the Debye-H\"{u}ckel equation with a boundary condition of prescribed surface charge,
\begin{equation}  \label{PB_DH}
  \nabla^2 \psi = \kappa^2 \psi,\quad \varepsilon\: \partial_y \psi = - q(x,0,z)
\end{equation}
The fluid flow satisfies Stokes' equations with an electrostatic body force
\begin{equation}  \label{Stokes}
  \eta \nabla^2 \textbf{\emph{u}} = \nabla p+\varepsilon \kappa^2 \psi E_t \hat{\textbf{\emph{x}}}, \quad \nabla \cdot \textbf{\emph{u}}=0,
\end{equation}
with the boundary conditions at $y=0$
\begin{equation}  \label{BCel}
  \textbf{\emph{u}}_t =  b(x,z) \partial_y \textbf{\emph{u}}_t, \quad  \hat{\textbf{\emph{y}}}\cdot \textbf{\emph{u}}=0,
\end{equation}
where   $\textbf{\emph{u}}_t = u \hat{\textbf{\emph{x}}} + w \hat{\textbf{\emph{z}}}$ is the lateral, and $ v =  \hat{\textbf{\emph{y}}}\cdot \textbf{\emph{u}}$ is normal to the surface velocities. We also neglect surface conduction (which tends to reduce EO flow) compared to bulk conduction ($Du \ll 1$)~\cite{epaps}. Far from the surface,  $y\rightarrow\infty$, $ \textbf{\emph{u}}$ approaches EO slip velocity  $\textbf{\emph{U}}=-\mathbf{M}\cdot \textbf{\emph{E}}_t$ and
\begin{equation}  \label{BCv}
   \psi\rightarrow 0, \quad \partial_y  \textbf{\emph{u}}\rightarrow 0.
\end{equation}
For a longitudinal configuration only velocity component parallel to $\textbf{\emph{E}}_t=E_t \hat{\textbf{\emph{x}}}$ remains. In case of transverse to applied field stripes, normal velocity $v \cdot \hat{\textbf{\emph{y}}}$ does not vanish due to mass conservation condition in (\ref{Stokes}), which can significantly modify the EO flow. Rigorous calculations~\cite{epaps} allow one to find exact solutions for $U^{\parallel,\perp}$, and thus obtain the eigenvalues of the EO mobility tensor:
\begin{equation}  \label{U_ort_thin}
   M^{\parallel, \perp} = M_1 \left(b^{\parallel, \perp}_{\rm{eff}}  \frac{q_2-q_1+q_2\kappa b}{q_1 b}+1\right),
\end{equation}
where the effective slip lengths are given by Eqs.(\ref{beff_par_largeH}),(\ref{beff_ort_largeH}).
The flow is thus anisotropic and there is a simple relationship between the EO mobility and hydrodynamic slip-length tensors~\cite{note2}
\begin{equation}\label{result}
    {\bf M} = M_1 \cdot\left[ \mathbf{I} +  \frac{{\bf b}_{\rm eff}}{b}\left(\frac{q_2}{q_1}\left(1 + \kappa b\right) - 1 \right)\right]
\end{equation}
 In the limit of $b/L \gg 1 $ the general expression transforms to Eq.~(\ref{squires}). When $b/L \ll 1$  we get isotropic EO flow
\begin{equation}\label{result_smallb}
    M =  \phi_1 {M}_1 + \phi_2 {M}_2
\end{equation}

\begin{figure}
\begin{center}
\includegraphics [width=8.5 cm]{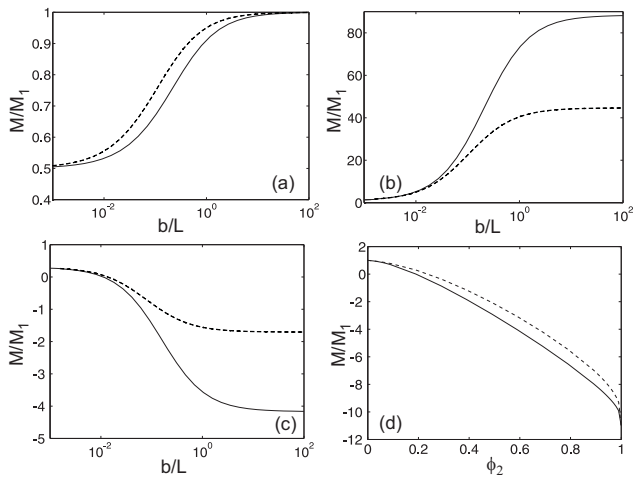}
\end{center}
  \caption{Eigenvalues  of   normalized EO mobility. Solid curves represent longitudinal, and dashed - transverse alignment of stripes with electric field. $M^{\parallel,\perp}/M_1$ vs amplitude of the local slip $b/L$ for (a) uncharged slip areas ($q_2=0, \phi_2=0.5, \kappa L = 10^2$), (b) uniform charge distribution ($q_2=q_1$; $\phi_2=0.45$, $\kappa L=10^3$) and (c) oppositely charged slip and no-slip areas ($q_2=-q_1$.  $\phi_2=0.35$, $\kappa L=10^2$). $M^{\parallel,\perp}/M_1$ plotted against (d) the fraction of gas sectors ($b/L=0.1$, $q_2=-q_1$, $\kappa L=10^2$). }
  \label{fig:electrokin}
\end{figure}

{\bf Discussion.}-- To demonstrate examples of very rich and unusual fluid behavior at the SH surface it is instructive to consider some limiting cases of Eq.~(\ref{result}) with different values of $q_1$ and $q_2$.
The results (shown in Fig.~\ref{fig:electrokin}) are somewhat remarkable. We see, in particular, that if gas area is uncharged, the EO flow related then only to the charge $q_1$ on the solid-liquid  interface is generally  \emph{inhibited} as compared with a homogeneous, solid no-slip surface with uniform charge density (see Fig.~\ref{fig:electrokin}(a))
\begin{equation}
{\bf M} = M_1 \cdot \left[ \mathbf{I} -  \frac{{\bf b}_{\rm eff}}{b}\right]
\label{cassie-formula-long}
\end{equation}
We remark and stress that in contrast to common expectations the situation described by Eq.(\ref{cassie-formula-long}) corresponds to $M^\parallel \le M^\perp$, i.e. the maximal directional mobility is attained in a transverse, and minimal -  in  longitudinal direction \cite{Mob_note}.
When $b/L \ll 1$  we simply get ${M} = \phi_1 {M}_1$. In other words, the (isotropic) EO mobility shows no manifestation of the slip, being equal to the surface averaged velocity generated by no-slip regions. This result coincides with expected for hydrophilic slip sectors. In the limit of $b/L \gg 1$ this inhibition becomes negligibly small, and we obtain the simple result of \cite{Squires08,bahga:2009}, where EO mobility becomes equal to ${M}_1$ regardless of the orientation or area fraction of the slipping stripes. These results suggest that although the absence of the screening cloud near the gas region tends to inhibit the effective EO slip, the hydrodynamic slip acts to suppress this inhibition.

The situation is very different if the slipping interface carries some net charge, which is not an unreasonable assumption~\cite{kirby2008}.
To gain some insight into the possible EO flow \emph{enhancement}, we consider first the case of uniform surface charge $q_1=q_2$, where Eq.~(\ref{result}) gives
\begin{equation}
{\bf M} = {M}_1 \cdot\left[ \mathbf{I} + \kappa {\bf b}_{\rm eff}\right]
 \label{const-q-formula}
\end{equation}
which might be seen as a natural tensorial analog of Eq.~(\ref{isotropic}).
  Fig.~\ref{fig:electrokin}(b) includes theoretical results calculated with Eq.~(\ref{const-q-formula}) for a geometry of stripes, and is intended to demonstrate that the flow is truly anisotropic and can exhibit a large enhancement from effective hydrodynamic slip, possibly by an order of magnitudes. We stress that such an enhancement is possible even at a relatively low gas fraction, i.e. when ${\bf b}_{\rm eff}$ is relatively small (but the amplification factor, $({\bf I} + \kappa {\bf b}_{\rm eff})$, might be huge). Also note that in this situation $M^\parallel \ge M^\perp$, i.e. the fastest/slowest direction can correspond only to longitudinal/transverse stripes.

An interesting scenario is expected for oppositely charged solid and gas sectors. If $q_1=-q_2$, then Eq.~(\ref{result}) transforms to
\begin{equation}
{\bf M} = {M}_1 \cdot\left[ \mathbf{I} - 2 \frac{{\bf b}_{\rm eff}}{b} - \kappa {\bf b}_{\rm eff}\right],
 \label{opposite-q-formula}
\end{equation}
which for $b/L \ll 1$ simply gives ${M} = {M}_1 [\phi_1 -\phi_2 (1 + \kappa b)]$. The calculation results for this situation are presented in Fig.~\ref{fig:electrokin}(c), and suggest a very rich fluid behavior. We see, in particular, that inhomogeneous  surface charge can induce EO flow along and opposite to the field, depending on the fraction of the gas area as shown in Fig.~\ref{fig:electrokin}(d). Already a very small fraction of the gas sectors would be enough to reverse the effective EO flow.  Another striking result is that electro-neutral surface ($ \phi_1 q_1 +\phi_2 q_2 = 0$) can generate extremely large EO slip. With our numerical example this corresponds to $\phi_2=0.5$. In other words, super-hydrophobic surface of average positive charge or even zero charge can induce an EO flow (different for longitudinal and transverse  to applied field stripes) in the direction of the applied field as if it is uniformly and negatively charged. These findings are similar to those of~\cite{anderson.jl:1989,ajdari2002} that the electrokinetic mobility depends on the charge distribution on the object, and not solely on its total charge. However, in our case the flow is dramatically amplified due to hydrodynamic effective slip.

\begin{figure}
\begin{center}
\includegraphics [width=8.5 cm]{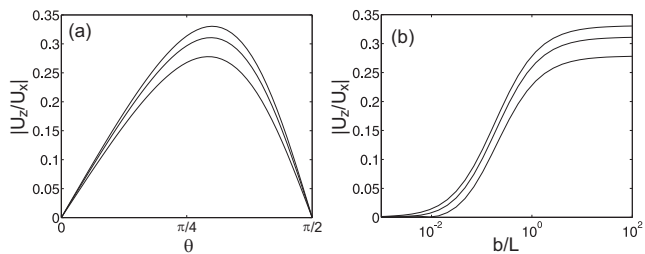}
\end{center}
  \caption{The ratio of EO velocity components, $|U_z/U_x|$, (at $\phi_2=0.5$, $\kappa L=10^2$) as a function of (a) angle $\theta$ at $b/L=10^2$  and (b) local slip length at the optimal angle $\theta$. From top to bottom $q_2/q_1 = 2, 1, 0.5$ }
  \label{fig:mix}
\end{figure}

These results may guide the design of SH surfaces for transverse electrokinetic flows in microfluidic devices~\cite{ajdari2002}. As we have shown above, effective EO mobility of anisotropic striped surfaces is generally tensorial, due to secondary flow transverse to
the direction of the applied electric field. Anisotropy ($|U_z/U_x|$) is maximized in certain direction $\theta_{\max}$ (as it is seen in Fig.\ref{fig:mix}a) and requires that $q_2/q_1$ and $\kappa b$ are as large as possible.
In a thick SH channel a transverse ``plug'' EO flow seems to be very fruitful direction compared to transverse hydrodynamic phenomena, where flow is ``twisted'' only near the wall~\cite{vinogradova.oi:2011}.

\begin{figure}
\begin{center}
\includegraphics [width=8.5 cm]{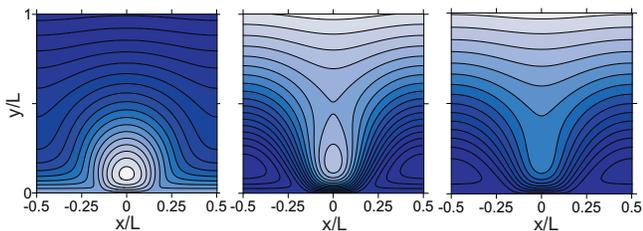}
\end{center}
  \caption{(Color online) Streamlines of the EO flow computed at $\phi_2=0.35$ and $\kappa L = 10^2$ for $q_2/q_1=-0.43$ and $\theta=\pi/2$. The origin of coordinates coincides with the center of the gas region. From left to right the local slip length is $b/L = 0.1$, $0.5$ and $10^2$.
  }
  \label{fig:vortex}
\end{figure}

Another mixing mechanism is related to the formation of patterns of steady convective rolls on the scale proportional to the texture period (Fig.~\ref{fig:vortex}). This can happen in the situation of oppositely charged and transverse to applied field stripes.
Fig. \ref{fig:vortex} illustrates the effect of the local slip on the morphology of  the steady rolls formation. We see, in particular, that increase of $b$ leads first to appearance of additional convective patterns near no-slip areas, and then to a transition to a flow morphology, where recirculation of a fluid is observed only at the no-slip regions. This, in turn, induces the flow reversal (as in Fig.~\ref{fig:electrokin}(d))

{\bf Concluding remarks.-- } We have described EO on inhomogeneously charged and slipping anisotropic surfaces. Our analysis provided the necessary tools to describe a significant modification of EO phenomena on SH surfaces: to quantify the inhibition and enhancement of flow, the transition from its anisotropy to isotropy, onsets of convective rolls formation and a relevant flow reversal, which can generate a huge EO slip even in the situation of a zero mean charge. Our results may find numerous applications in microfluidic lab-on-a-chip devices.

\bibliographystyle{rsc}
\bibliography{electroosmos}

\begin{thebibliography}{10}
\bibitem{Mob_note}
This surprising observation likely reflects the fact, that the reduction of the
  interfacial stress (and the mobility) due to slip in the absense of the
  diffuse layer should be smaller for transverse stripes since it is partly
  compensated by the local pressure drop.
\end{thebibliography}

\end{document}


\title{\textbf{Supplementary information.\\ \emph{Electro-osmosis on anisotropic super-hydrophobic surfaces}}}

\author{Aleksey V. Belyaev}

\author{Olga I. Vinogradova}


\maketitle

\section{Dukhin and P\'{e}clet numbers in the presence of slip}

Surface conduction in presence of slip is characterized by full Dukhin number, which is given by~\cite{khair_squires2009}:
\begin{equation}  \label{Du}
   Du = \frac{4(1+m)}{\kappa L}\sinh^2\left(\frac{ze\zeta}{2k_B T}\right)+\frac{2mb}{L}\sinh^2\left(\frac{ze\zeta}{4k_B T}\right),
\end{equation}
where $m=2\varepsilon(k_B T/ze)^2/(\eta D)$, and $D$ is the ion diffusivity, which is assumed equal for both types of ions.
The first term in (\ref{Du}) is the Dukhin number for the no-slip surface, $Du_{b=0}$, while the second one, $Du_{b}$,  is due to hydrodynamic slip. In the Debye-H\"{u}ckel model
\begin{equation}  \label{DH_limit}
   \frac{ze\zeta}{k_B T}\ll 1.
\end{equation}
This restriction (\ref{DH_limit}) allows the simplification:
\begin{equation}  \label{Du_ns}
   Du_{b=0}\approx \frac{4(1+m)}{\kappa L}\left(\frac{ze\zeta}{2k_B T}\right)^2,
\end{equation}
\begin{equation}  \label{Du_s}
   Du_{b}\approx  \frac{2mb}{L}\left(\frac{ze\zeta}{4k_B T}\right)^2.
\end{equation}
The parameter  $m\approx 100/z^2$,  and $\kappa L \gg 1$  since EDL is thin. Whence $Du_{b=0}\approx 1/(z^2 \kappa L)\ll 1$ provided the potential is low.  For a ``slippery part'' of Du we evaluate $ Du_b\approx 0.5 b/L$  if  $e\zeta/(k_B T)\approx 0.1$, and $Du_b\approx 5\cdot 10^{-3} b/L$  for  $e\zeta/(k_B T)\approx 0.01$. Therefore for increasing surface charge (potential) and $b/L$, the conductivity of the diffuse layer can become
comparable to the bulk, and surface condition must be considered.

The P\'{e}clet number, $Pe=UL/D$ ,  in presence of slip can be evaluated as
\begin{equation}  \label{Pe}
   Pe= \frac{q_2 E_t(1+b\kappa)L}{\kappa\eta D}.
\end{equation}
Typically, electroosmotic velocity is of order is
of order micrometers per second for no-slip surfaces~\cite{anderson.jl:1989}. For nano-scale patterns $L<1$ $\mu$m and typical ion diffusivities $D\approx 10^{-6}$ ${\rm cm}^2/{\rm s}$  this gives $Pe_{b=0}<0.01 \ll 1 $. The slip implies a correction factor $(1+b\kappa)$ , which suggests that the convective ion transport can safely be neglected only for $b\kappa < 10$. Larger values of $b\kappa$ should relax this standard approximation of small $Pe$.

\section{Electro-osmotic velocity in eigendirections}

{\bf Longitudinal stripes.}-- In this configuration only $x-$velocity component remains, and the Stokes equation takes the form
\begin{equation}  \label{Stokes_par}
  \left(\partial^2_y+\partial^2_z\right)u=\varepsilon \kappa^2 \psi E_t
\end{equation}
We expand surface charge density in a Fourier series, and the potential is then
\begin{equation}  \label{psi_par}
  \psi(y,z) = \frac{\langle q\rangle}{\varepsilon\kappa}e^{-\kappa y} + \sum\limits_{n=1}^\infty \frac{q_n}{\varepsilon \xi_n}e^{-\xi_n y} \cos{\lambda_n z},
\end{equation}
where $\xi_n=\sqrt{\kappa^2+\lambda_n^2}$, $\lambda_n = 2 n \pi /L$ , $\langle q \rangle = q_1 \phi_1 + q_2 \phi_2 $ is the mean surface charge, and
\begin{equation}
  q_n=\frac{2(q_2-q_1)}{\pi n}\sin{\frac{\pi n\delta}{L}}.
\end{equation}
The general solution to (\ref{Stokes_par}) for $u(y,z)$  has the form
\begin{equation}  \label{upar1}
  u(y,z) =  U^{\parallel} + \sum\limits_{n=1}^\infty U_n e^{-\lambda_n y} \cos{\lambda_n z} + \frac{\varepsilon E_t}{\eta}\psi,
\end{equation}
where $U_n$ and $U^{\parallel}$ are determined by the slip boundary conditions. Imposing them on (\ref{upar1}) in a \emph{thin EDL} limit yields  a dual series
\begin{multline}  \label{dual_series_par_thina}
   U^{\parallel}
   + \sum\limits_{n=1}^\infty U_n(1+b\lambda_n)\cos{\lambda_n z} = {} \\
    {} = -\frac{E_t q_2}{\kappa\eta}(1+\kappa b), \quad |z|\leq \delta/2,
\end{multline}
\begin{equation}  \label{dual_series_par_thinb}
   U^{\parallel}
   + \sum\limits_{n=1}^\infty U_n\cos{\lambda_n z}= - \frac{E_t q_1}{\kappa\eta}, \quad \delta/2<|z|\leq L/2,
\end{equation}
which can be solved exactly by using a technique~\cite{belyaev.av:2010a} to obtain the thin-EDL electro-osmotic  velocity:
\begin{equation}  \label{U_par_thin}
   U^{\parallel} = - \frac{E_0}{\eta} \left(b^\parallel_{\rm{eff}} \frac{q_2-q_1+q_2\kappa b}{\kappa b}+\frac{q_1}{\kappa}\right).
\end{equation}

{\bf Transverse stripes.}--Although an external pressure gradient is equal to zero, local pressure variations contribute into a non-zero term $\nabla p $ in the Stokes equation, so that the flow is essentially two-dimensional. We first introduce a stream function $f(x,y)$
\begin{equation}  \label{Ff}
   \partial_y f = u, \quad \partial_x f = -v,
\end{equation}
which obeys inhomogeneous biharmonic equation:
\begin{equation}  \label{Eq_Ff}
   \nabla^2 \nabla^2 f = E_t \frac{\varepsilon\kappa^2}{\eta}\frac{\partial\psi}{\partial y}
\end{equation}
Here $u$ and $v$ are $x$ and $y$ velocity components, correspondingly. The general periodic solution to (\ref{Eq_Ff}) has the form
\begin{multline}  \label{Ff_sol}
   f(x,y)=U^\perp y+\sum\limits_{n=1}^\infty{\left(\frac{E_t q_n}{\kappa^2 \eta}+g_n y\right)e^{-\lambda_n y}\cos{\lambda_n x}} + {} \\
   {} +\frac{E_t \varepsilon}{\kappa^2 \eta}\frac{\partial\psi}{\partial y}
\end{multline}
Here the potential $\psi(x,y)$ has exactly the form ($\ref{psi_par}$) with $z$ replaced by $x$.
The dual series problem in a \emph{thin EDL} limit can be written as
\begin{multline}  \label{dual_series_ort1a}
   U^{\perp}
   + \sum\limits_{n=1}^\infty a_n(1+2b\lambda_n)\cos{\lambda_n x} = {} \\
    {}= -\frac{E_t q_2}{\kappa\eta}(1+\kappa b), \quad |x|\leq \delta/2,
\end{multline}
\begin{equation}  \label{dual_series_ort1b}
   U^{\perp}
   + \sum\limits_{n=1}^\infty a_n\cos{\lambda_n x}= - \frac{E_t q_1}{\kappa\eta}, \quad \delta/2<|x|\leq L/2,
\end{equation}
where
\[
  a_n=g_n+\frac{E_t q_n}{\eta\kappa^2}(\xi_n-\lambda_n).
\]

These dual series can be solved exactly to obtain
\begin{equation}  \label{U_ort_thin}
   U^{\perp} = - \frac{E_0}{\eta} \left(b^\perp_{\rm{eff}} \frac{q_2-q_1+q_2\kappa b}{\kappa b}+\frac{q_1}{\kappa}\right)
\end{equation}

We emphasize that a comparison of Eqs.(\ref{U_par_thin}) and (\ref{U_ort_thin}) indicates that the EO flow is generally anisotropic, so that our results do not support an earlier conclusion~\cite{bahga:2009} that the electro-osmotic mobility tensor is isotropic in the thin EDL limit. This inconsistensy~\cite{bahga:2009} (due to an erroneous expression for a transverse electro-osmotic velocity, where factor of 2 was lost) has been corrected for a case $b_2=\infty$ in~\cite{vinogradova.oi:2011}.

\bibliographystyle{rsc}
\bibliography{electroosmos}